\documentclass[11pt]{article}
\usepackage{hyperref,xcolor}
\usepackage{cite}
\usepackage{amsmath}
\usepackage{graphicx, tikz}
\usepackage{amssymb}
\usepackage{footnote}
\makesavenoteenv{minipage}
\newcommand\bra[2][]{#1\langle {#2} #1\rvert}
\newcommand\ket[2][]{#1\lvert {#2} #1\rangle}

\newcommand{\mc}{\mathcal}

\renewcommand{\title}[1]{%
    \bigskip%
    \begin{center}%
    \Large\bf #1%
    \end{center}%
    \vskip .2in}

\renewcommand{\author}[1]{%
    {\begin{center}
    #1
    \end{center}}}
\newcommand{\address}[1]{\vspace{-1.7em}\vspace{0pt}
    {\begin{center}
    \it #1
    \end{center}}}

\begin{document}

\begin{titlepage}
\title{$\mc{PT}$-symmetry of Particle mixing theories and the equation of motion matrix  }

\author
{
	Kawaljeet Kaur $\,^{\rm a,b}$,
Biswajit Paul  $\,^{\rm a,c}$,

}
\address{$^{\rm a}$ National Institute of Technology Agartala \\
 Jirania, Tripura -799 055, India }

\footnote{
{$^{\rm b}$\tt kakawaljeet@gmail.com,}
 
{$^{\rm c}$\tt biswajit.thep@gmail.com}}
\begin{abstract}
 A non-Hermitian complex scalar field model is considered from its $\mc{PT}$ symmetric aspect. A matrix constructed from the Euler-Lagrange equations of motion is utilized  to analyze the states of the model. The model has two mass terms which determine the real or complex nature of the eigen values. A mismatch is found in the Lagrange equations of motion of the  fields as the equations  do not agree with the other after complex conjugation of the either. This is resolved by exploiting a preferred similarity transformation of the Lagrangian. The discrepancy even at the Hamiltonian level is found to have vanished once we consider the similarity transformed Hamiltonian. 

\end{abstract}

\end{titlepage} 
\section{Introduction}
 Every physical theory involving  natural phenomenon is governed by some kind of symmetry. A very important symmetry in  quantum field theory  models is the parity $\mc{P}$ and time reversal $\mc{T}-$symmetry. $\mc{PT}$ -symmetric theories are invariant under the space and time reflections. The concept of $\mc{PT}$ symmetry is not new in physics as it is frequently used in various  branches of physics like acoustics \cite{zhu}, photonics \cite{morales, wilkey, longhi}, electronics \cite{yang}, plasmonics \cite{luhu} and many more. Recently,   it was revived by the works of Bender et al due to their interesting works on non-Hermitian Hamiltonians \cite{bender1998, bender2007}. It was discovered that apart from the $\mc{P}$ and the $\mc{T}$ operators these theories also poses a new operator, called the $\mc{C}$ operator which commutes with the $\mc{PT}$ and the Hamiltonian both. These three constitute a $\mc{CPT}$ symmetry which is unlike the charge, parity and time reversal symmetry appearing usually in quantum field theories. Apart from these, the $\mc{PT}$ symmetric theories can have antilinear symmetry as shown in \cite{mannheim2018}. As the eigen spectrum for  non-Hermitian Hamiltonians  is incomplete, the Hamiltonians form  a non-diagonalizable Jordan block form which is used to study the antilinearity. In fact, Mannheim et al has shown that antilinearity is a more general requirement for the reality of the energy eigenvalues \cite{mannheim2018}.

Non-Hermitian Hamiltonians appear in various context of physics  \cite{mannheim2016, alexandre2015, alexandre2015I, bender2012, bender2005}. Earlier the non-Hermitian models were thought to give unbounded energy eigen values and were not a fit candidate as  quantum mechanical models. However, it was shown that owing to the nature of $\mc{PT}$ symmetry, these Hamiltonians can have bounded from below energy levels \cite{bender1998, paul2020}. These interesting results led to reinvestigation of the non-Hermitian Hamiltonians from $\mc{PT}$ symmetric point of view.   

An interesting example of non-Hermitian model we can take by considering  complex scalar fields with anti-Hermitian mass term as studied in \cite{alexandre2017} which actually is a simple  non-Hermitian free scalar field model(without tadpole). It was shown in \cite{alexandre2017} that the owing to the non-Hermitian term, the Euler-Lagrange equations of motion shows discrepancy when one takes the complex conjugate of the equations of motion. However, Mannheim in \cite{mannheim2019} has considered a different approach by applying a similarity transformed Hamiltonian. When one calculate the equations of motion using this similarity transformed Lagrangian or Hamiltonian, it is found that the discrepancy is resolved and thus the both the equations of motion agree with their complex conjugate counter part. The proper operator for the similarity transformation was found out by exponentiating the product of the field and it's corresponding momenta. 

 The non-Hermitian complex scalar field model was considered by Jean et al   consisting of a (squared) mass matrix. The eigenvalues and eigen vectors of this mass matrix were used to form the basis vectors of the corresponding quantum theory\cite{alexandre2017}. On the other hand, the equations of motion form a matrix as shown in \cite{mannheim2019} and this can be very important for analyzing the quantum theory of the model. The eigenvalues and eigenvectors of the EoM matrix are used to form the corresponding normalized eigenvectors of the model.

 
 In this paper, we have considered a complex scalar field model with non-Hermitian terms that consist of mass mixing. Instead of the mass matrix arising from the Lagrangian, we have considered the matrix formed from the equations of motion(EoM). The equations of motion of the fields form a matrix and that is utilised to analyse the model from $\mc{PT}$-symmetric point of view. A similarity transformation is used as in \cite{mannheim2019} to form the operator $S$. Using this operator $S$ the $V$-operator is constructed. This $V-$operator is actually related to the antilinear symmetry of the model and was mentioned in \cite{mannheim2018}. Another aspect we have considered here is the discrepancy in the equations of motion of the two conjugate fields for this model. The Euler-Lagrange equations of motion for the field and the conjugate fields has a difference of negative sign once we consider the complex conjugation of the either. To get rid of this we first perform a similarity transformation of the Lagrangian and found that now the mismatch in the equations of motion is gone. Even the Hamiltonian equations of motion calculated after the suitable similarity transformation agree for both the conjugate fields. So, it seems that the equations of motion matrix is a more viable option than just considering the mass matrix of the system.

 
 The structure of this paper is as follows. In Section 2 we have considered the anti-Hermitian model with mass mixing and its symmetries under $\mc{P}$ and $\mc{T}$. Section 3 deals with the equations of motion of the fields and application of Mass matrix for construction of the left and right eigen vectors. In Section 4 we have considered the mismatch   in the equations of motion for the conjugated fields. A way to resolve this mismatch has been shown. Finally, we conclude with Section 5. 
\section{Non-Hermitian mass mixing model }
We consider the general transformation of the scalar fields under parity($\mc{P}$) and time reversal($\mc{T}$) as  

\begin{eqnarray} 
	\mc{P}\phi(t,\vec{x})\mc{P}^{-1} = \phi'(t,-\vec{x}) \\ 
	\mc{T}\phi(t,\vec{x})\mc{T}^{-1} = \phi^*(-t,\vec{x}). 
\end{eqnarray} 

In addition to these two operators, the unbroken $\mc{PT}$-symmetric theories also have another more physically motivated symmetry operator, called the $\mc{C}$-operator as put forward by Bender et al. \cite{bender2004}. Generally, the $\mc{C}$ operator is constructed by considering the sum over all the states although there exist other ways like perturbative methods \cite{bender2004I}. 

We consider that the field $\phi$ consists of the doublet $\phi_1$ and $\phi_{2}$ of which one behaves as scalar and the other as pseudoscalar under $\mc{P}$ and $\mc{T}$. This is due to the complex nature and both the fields manifestly have an interpretation as a coupled source and sink with gain and loss nature. A non-Hermitian model with this kind of field combination can be considered from \cite{alexandre2020} where the Lagrangian is given by 

\begin{eqnarray} 
	\mc{L} = \partial_\nu \phi_1^* \partial^\nu \phi_1 + \partial_\nu \phi_2^* \partial^\nu \phi_2 - (m_1^2 |\phi_1|^2 + m_2^2 |\phi_2|^2) - \mu^2(\phi_1^* \phi_2 - \phi_2^* \phi_1). 
	\label{lag1} 
\end{eqnarray} 

In this Lagrangian the fields are $\{ \phi_1, \phi_2, \phi^*_1, \phi^*_2\}$. Clearly, the Lagrangian is not Hermitian i.e. $\mc{L} \ne \mc{L}^\dagger$. The true non-Hermiticity is evident when we consider invariance under the CPT transformations as discussed in \cite{mannheim2018, mannheim2019}. To understand we consider the following $\mc{CPT}$ transformation of the field

\begin{eqnarray} 
	\phi_1(x_\mu) \rightarrow \phi_1^*(-x_\mu), \ \ \ \ \ \phi_2(x_\mu) \rightarrow -\phi_2^*(-x_\mu)  
\end{eqnarray} 





Therefore, the model also has anti-linear symmetry \cite{mannheim2018}. This antilinear symmetry is required for the time independence of the inner products and reality of the energy eigen values.

Now, consider the  $U(1)$ gauge transformations  

\begin{eqnarray} 
	&& \phi_1(x) \rightarrow \phi'_1(x) = \exp(i\alpha) \phi_1(x), \\ 
	&& \phi_2(x) \rightarrow \phi'_2(x) = \exp(-i\alpha) \phi_2(x). 
	\label{symm1} 
\end{eqnarray} 

Where $\alpha$  corresponds to the symmetry parameter. It is easy to notice that under these transformations the Lagrangian (\ref{lag1})  is not invariant i.e $\mc{L} \ne \alpha \mc{L}$. This unusual behavior is due to the non-Hermiticity of the model. If we consider a different gauge transformation

\begin{eqnarray} 
	&& \phi_1(x) \rightarrow \phi'_1(x) = \exp(i\alpha) \phi_1(x), \\ 
	&& \phi_2(x) \rightarrow \phi'_2(x) = \exp(i\alpha) \phi_2(x), 
	\label{symm2} 
\end{eqnarray} 

we see that the Lagrangian is invariant. The symmetry (\ref{symm1}) leads to a conserved current but is not a symmetry of the Lagrangian (\ref{lag1}) whereas (\ref{symm2}) is a symmetry of the Lagrangian but does not lead to a conserved current. This piculiarity of the Lagrangian was discussed in detail by Alexander et al and alternative variational procedure was discussed in  \cite{alexandre2017}.

Therefore, the Lagrangian is not invariant under the global $U(1)$ transformation due to the complex nature. 
\section{The equation of motion matrix}
Decomposing the fields into their complex combination form as
\begin{eqnarray}
	\phi_1 = \frac{1}{\sqrt{2}}(X_1 + iX_2), \ \ \ \  \phi_2 = \frac{1}{\sqrt{2}}(Y_1 - iY_2)
\end{eqnarray}

the Lagrangian (\ref{lag1}) becomes \cite{mannheim2019}
\begin{eqnarray}
	\nonumber 
	I  &=& \int d^4x \Big(\partial_\nu X_1 \partial^\nu X_1 + \partial_\nu X_2 \partial^\nu X_2 + \partial_\nu Y_1 \partial^\nu Y_1 + \partial_\nu Y_2 \partial^\nu Y_2  \\
	 && - \frac{1}{2}(m_1^2( X_1^2 + X_2^2) + m_2^2 (Y_1^2 + Y_2^2)) - i \mu^2  (X_1 Y_2 - X_2 Y_1) \Big).
	\label{action1}
\end{eqnarray}
Varying the action (\ref{action1}) we obtain the Euler-Lagrange equations of motion corresponding to the fields which are given by
\begin{eqnarray}
	-\Box X_1 &=& m_1^2 X_1 + i \mu^2 Y_2, \\ \label{lag_emo1}
	-\Box X_2 &=& m_1^2 X_2 - i \mu^2 Y_1, \\ \label{lag_emo2}
	-\Box Y_1 &=& m_2^2 Y_1 - i \mu^2 X_2, \\ \label{lag_emo3}
	-\Box Y_2 &=& m_2^2 Y_2 + i \mu^2 X_1. \\	\label{lag_emo4}
\end{eqnarray}
 Minimum of these can be obtained by equating the RHS of the above equations to zero. The solutions are
\begin{eqnarray}
	\bar{Y}_1= \frac{i \mu^2 \bar{X}_2}{m_2^2}, \bar{Y}_2 = - \frac{i \mu^2 \bar{X}_2}{m_2^2}, 
\end{eqnarray}
The equations of motion   corresponding to the fields form a  matrix  given by 
\begin{center}
 $\begin{pmatrix}
	
	-\Box X_1 \\ -\Box Y_2 \\ -\Box X_2 \\ -\Box Y_1
\end{pmatrix}$ = $\begin{pmatrix}
	m_1^2 & i\mu^2 & 0 & 0\\ 
	i \mu^2 & m_2^2 & 0 & 0 \\
	0 & 0  &  m_1^2  & -i \mu^2 \\
	0 & 0 & -i \mu^2 & m_2^2 
\end{pmatrix}$  $\begin{pmatrix}
	X_1 \\  Y_2 \\  X_2 \\  Y_1
\end{pmatrix}$.	
\end{center}
 
We shall identify the matrix formed by the above equation as 
\begin{equation}
	M = \begin{pmatrix}
		m_1^2 & i\mu^2 & 0 & 0\\ 
		i \mu^2 & m_2^2 & 0 & 0 \\
		0 & 0  &  m_1^2  & -i \mu^2 \\
		0 & 0 & -i \mu^2 & m_2^2 
	\end{pmatrix}. \label{eom_matrix}
\end{equation}
Our main concentration for analysis of the model will be this matrix $M$. The equation of motion matrix (\ref{eom_matrix}) M has  eigenvalues which are given by

\begin{equation}
	\lambda_{\pm} = \frac{1}{2}\Big((m_1^2 + m_2^2) \pm \sqrt{(m_1^2 - m_2^2)^2 - 4 \mu^4}\Big). \label{eigen_values}
\end{equation}

There may be interesting sectors as the eigenvalues can be real or imaginary for $(m_1^2 - m_2^2)^2$ greater than, less than or equal to $  4\mu^4$. The theory being a non-Hermitian one $H \ne H^\dagger$ with unbroken $\mc{PT}$-symmetry, we can find an operator $V$ that will after a similarity transformation  give 
\begin{equation}
	VHV^{-1} = H^\dagger.
\end{equation}
The existence of the V-operator is required in the non-Hermitian theories for the reason that the eigen states corresponding to the non-Hermitian Hamiltonian do not have left-right symmetry. Meaning that we require two different states namely left eigen state $\ket{R_{\pm}}$ and right eigen state $\bra{L_{\pm}}$. The above condition also ensures that the inner products defined using these V-based operators are time independent \cite{mannheim2013}.
	
To get the $V$ and $S$ operators we notice that the matrix M can be written in distinct 2D blocks given by
 
 \begin{center}
 $ N =  \begin{pmatrix}
 	C+A & i B \\  i B &  C - A
 \end{pmatrix}$ 
\end{center}
so that $C+ A = m_1^2$, $C - A = m_2^2$ and $\lambda_{\pm} = C \pm \sqrt{A^2-B^2}$. In order to check the completeness of the eigenspectrum we must built the matrix S as we already have seen that the Hamiltonian is not Hermitian $H \ne H^\dagger$. While the $S$-matrix is given by

\begin{equation}
	S = \frac{1}{2 (A^2 - B^2)^{\frac{1}{4}}}\begin{pmatrix}
		\sqrt{A+B} + \sqrt{A-B} & i (\sqrt{A+B} - \sqrt{A-B}) \\  i (\sqrt{A+B} - \sqrt{A-B}) &  \sqrt{A+B} + \sqrt{A-B}
	\end{pmatrix}.
\end{equation}
The V-matrix thus can be obtained from \cite{mannheim2018}
\begin{equation}
	V = S^\dagger S = \frac{1}{(A^2 - B^2)^{1/2}} \begin{pmatrix}
		A & iB \\ -i B & A
	\end{pmatrix}.
\end{equation}
We now form the eigen vectors of the matrix M. As discussed above that the theory requires two vectors viz. right and left eigen vetors we calculate them for  (\ref{eom_matrix}) as \cite{mannheim2019}
\begin{equation}
	L_0 = \frac{1}{\sqrt{2} [(m_1^2 - m_2^2)^2 - 4 \mu^4]^{1/4}} \begin{pmatrix}
		(A+B)^{1/2} + (A-B)^{1/2}, & i ((A+B)^{1/2} - (A-B)^{1/2})
	\end{pmatrix}
\end{equation}
and 

\begin{equation}
	R_0 = \frac{1}{\sqrt{2} [(m_1^2 - m_2^2)^2 - 4 \mu^4]^{1/4}} \begin{pmatrix}
		& (A+B)^{1/2} + (A-B)^{1/2}, \\ & i ((A+B)^{1/2} - (A-B)^{1/2})
	\end{pmatrix}.
\end{equation}
The above vectors follow the relations
\begin{eqnarray}
	L_0 R_0 =1, L_1 R_1 =1, L_0 R_1 = L_1 R_0 =0.
\end{eqnarray}
Thus $L_0$ and $R_0$ forms a complete sets of eigenvectors.

\subsection{Similarity transformation operator}
Now, we write down the Hamiltonian. The Hamiltonian from the action (\ref{action1}) is geven by
\begin{equation}
	H = \sum_{i=1}^4 \Big(\frac{1}{2}P_i^2 - \frac{1}{2} (\nabla X_i)^2 + \frac{1}{2} m_i^2 X_i^2 \Big)+ \mu^2  i (X_1Y_2-X_2Y_1) 
\end{equation}.
Where $P_i$ is the momenta corresponding to $X_i$. 

A different approach for explaining this discrepancy was explained by Mannheim \cite{mannheim2019}. It was shown that both the actions are connected by a similarity transformation. For the present model under consideration, we take the operator for similarity transformation as 
\begin{equation}
	\bar{S} = \exp\Big(-\theta\sum_{i=1,2} P_i X_i \Big). \label{sOperator}
\end{equation}
The similarity transformation of the Hamiltonian is given by 
\begin{equation}
	\bar{S} H \bar{S}^{-1} = \sum_{i=1,2} \Big(\frac{1}{2}P_i^2 e^{2i\theta} - \frac{1}{2} (\nabla X_i)^2 + \frac{1}{2} m_i^2 X_i^2 e^{-2i\theta} \Big)+ \mu^2  i (X_1Y_2-X_2Y_1) e^{-i \theta}.
	\end{equation}
A choice of $\theta = \pi/2$ gives
\begin{equation}
	\bar{S}H\bar{S}^{-1} = H^\dagger.
\end{equation}
The basic fields thus transforms under the action of $\bar{S}$ as
\begin{eqnarray}
	\bar{S} X_i \bar{S}^{-1} &=& -iX_i,  \\	 
	\bar{S} \Pi_i \bar{S}^{-1} &=& i \Pi_i . 
\end{eqnarray}

Also, the fields change as
\begin{eqnarray}
	\bar{S} \phi_{i} \bar{S}^{-1} = -i\phi_{i}, \ \ \ \ \ \bar{S} \phi_{i}^* \bar{S}^{-1} = -i\phi_{i}^*.
\end{eqnarray}
\section{Discrepencies in the equations of mtion}
Now we consider the equations of motion of the model. For that purpose, we consider the action of the Lagrangian density (\ref{lag1}) and vary it to find the Euler-Lagrange equations of motions as done in  \cite{alexandre2017}. However, there appears ambiguity of a '-' sign in one of the term while we vary the action with respect to $\phi_1$ and $\phi_1^*$. That is, one equation is not the complex conjugate to the other as expected. This issue was resolved by taking the complex conjugation of the action in the first place and then varying it with respect to $\phi_1^*$. This anomalous phenomenon happened due to the complex nature of the model.  

We consider the action (\ref{lag1}) where the fields are ($\phi_1, \phi_2, \phi^*_1, \phi^*_2 $). The Lagrangian equations of motion for the fields $\phi_1, \phi_2, \phi^*_1, \phi^*_2 $ are respectively given by 
\begin{eqnarray}
	(\Box + m_1^2)\phi_1^* - \mu^2 \phi_2^* &=& 0,  \label{eom_phi_1}\\ 
	(\Box + m_2^2)\phi_2^* + \mu^2 \phi_1^* &=& 0, \label{eom_phi_2}\\ 
	(\Box + m_1^2)\phi_1 + \mu^2 \phi_2 &=& 0, \label{eom_phi_3}\\ 
	(\Box + m_2^2)\phi_2 - \mu^2 \phi_1 &=& 0 \label{eom_phi_4}.
\end{eqnarray}  
It is expected by usual notion that the pair of equations \ref{eom_phi_1} and (\ref{eom_phi_2}) should be complex conjugate of (\ref{eom_phi_3}) and (\ref{eom_phi_4}) respectively. But it can be seen that they are not. There is a discrepancy of a negative sign in the terms containing $\mu^2$. This was observed in \cite{alexandre2017} and the issue was resolved by claiming a non-trivial definition of variational principle. Rather, the authors have proposed a modified variational scheme. This, however, was addressed subsequently also by Mannheim \cite{mannheim2019}. In order to solve this mismatch of equations of motion, the author has proposed a similarity transformed action given in (\ref{sOperator}). The similarity transformed Lagrangian density is given by 

\begin{eqnarray}
	\mathcal{L}' = \bar{S} \mc{L} \bar{S}^{-1} =     -\partial_\nu \phi_1 \partial^\nu \phi_1^* - \partial_\nu \phi_2 \partial^\nu \phi_2^* + m_1^2 \phi_1 \phi_1^* + m_2^2 \phi_2 \phi_2^* +  i \mu^2 ( \phi_1^* \phi_2 - \phi_2^* \phi_1 ).    \label{lag2}
\end{eqnarray}
 We have calculated the Euler-Lagrange equations of motion for the above Lagrangian (\ref{lag2}) corresponding to $\phi_1, \phi_2$ which are complex conjugate of the equations of motion for $\phi_1^*, \phi_2^*$. Hence the discrepency of equations of motion is resolved. 
 However, one may argue about the Hamiltonian equations of motion. The Hamiltonian corresponding to the action (\ref{lag2}) is given by 
 \begin{equation}
 	H = -\Pi_1^*\Pi_1 + \Pi_2^*\Pi_2 - \partial^i \phi_1^* \partial_i\phi_1 - \partial^i \phi_2^* \partial_i\phi_2 - m_1^2 \phi_1^* \phi_1 + m_2^2 \phi_2 \phi_2^* - i\mu^2 (\phi_1^* \phi_2 - \phi_2^* \phi_1).
 \end{equation} 
The Hamiltonian equations of motion are calculated as usual and found to be
\begin{eqnarray}
\dot{\phi}_1 &=& -\Pi_1^*, \\ 
\dot{\phi}_1^* &=& -\Pi_1^, \\
\dot{\phi}_2 &=& -\Pi_2^* , \\ 
\dot{\phi}_2^* &=& -\Pi_2 , \\
\dot{\Pi}_1 &=& -\partial_i \partial^i\phi_{1}^* + m_1^2\phi_1^* - i\mu^2\phi_2^*, \\ 
\dot{\Pi}_1^* &=& -\partial_i \partial^i\phi_{1} + m_1^2\phi_1 + i\mu^2\phi_2 ,\\ 
\dot{\Pi}_2 &=& -\partial_i \partial^i\phi_{2}^* - m_2^2\phi_2^* + i\mu^2\phi_1^*, \\ 
\dot{\Pi}_2^* &=& -\partial_i \partial^i\phi_{2}^* - m_2^2\phi_2 - i\mu^2\phi_1^* .\\ 
\end{eqnarray} 
Here also the equations of motion agree for both the cases of complex conjugates.

 \section{Conclusion}
 $\mc{PT}$-symmetric non-Hermitian theories have become a very interesting topic among physicists in recent years. Despite various non-trivial results, they are still important in the context of physics. In this paper we have considered a $\mc{PT}$-symmetric complex scalar field model with non-Hermitian mass terms. The equations of motion for this model form a matrix and this matrix has been utilized for the further analysis of the quantum aspects of the model. The eigen values of the EoM matrix are found to be real or imaginary depending on the pair of mass. Thus, the eigen vectors and their characteristics also change depending on the masses. The equation of motion matrix has been written in a Jordan block form which is utilized for construction of the $V$ operator that exist due to the anilinear symmetry of the model. The left and right eigen vectors form a complete basis.

 Another very important aspect we have seen for these kinds of models is the complex conjugation of the equations of motion. The equations of motions corresponding to the  field and its conjugate are not conplex conjugate of one another. There is a mismatch of negative sign. This issue we have solved by considering a similarity transformation. Now the EoM for both the conjugated fields agree after complex conjugation of one another. Not only this, if we consider the Hamiltonian of this similarity transformed Hamiltonian, there appears no mismatch.

\end{document}